# Apparent Dresselhaus coefficient in (001) GaAs quantum wells: Correlation-time renormalization in D'yakonov–Perel' spin relaxation

Yuzo Ohno[1,2], Jun Ishihara[3], and Satoshi Iba[2]

1. Institute of Pure and Applied Sciences, University of Tsukuba, 1-1-1 Tennoudai, Tsukuba, 305-8573 Japan
2. Research Institute for Hybrid Functional Integration, National Institute of Advanced Industrial Science and Technology (AIST), Umezono 1-1-1, Central 2, Tsukuba, Ibaraki 305-8568, Japan
3. Department of Materials Science, Graduate School of Engineering, Tohoku University, Sendai, 980-8579, Japan

[†]Corresponding authors: Correspondence to Yuzo OHNO (ono.yuzo.gb@u.tsukuba.ac.jp) and Satoshi IBA (s.iba@aist.go.jp)

**Abstract**

We study the Dresselhaus coefficient inferred from D'yakonov–Perel' spin relaxation in bulk GaAs and (001)GaAs quantum wells using nonballistic Monte Carlo simulations. In bulk GaAs, inelastic LO-phonon simulations reproduce the spin relaxation with a cubic Dresselhaus coefficient $\gamma_{\rm 3D} \simeq 12.4$ eVÅ$^3$ and a cubic-field correlation time $\tau_3^{\rm 3D} \simeq 150$ fs. In a projected two-dimensional quantum-well model, however, $k_z^2$ is replaced by the static expectation value $\langle k_z^2 \rangle$, converting the dominant Dresselhaus field into a term linear in the in-plane wave vector. We show that this projection changes the D'yakonov–Perel' correlation kernel from $K_3^{\rm 2D}$ to a first- and third-order hybridized kernel $K_{1+3}^{\rm 2D}$, for which the correlation time $\tau_{1+3}^{\rm 2D}$ approaches 235 fs in the strict two-dimensional limit, in contrast to the third-order correlation time $\tau_3^{\rm 2D}$ of 130 fs. Consequently, the coefficient entering the projected two-dimensional model is not the intrinsic cubic coefficient itself but an apparent coefficient, $\gamma_{\rm 2D}^{\rm app} = \gamma_{\rm 3D}\sqrt{\tau_3^{\rm 2D}/\tau_{1+3}^{\rm 2D}}$. Since $\tau_{1+3}^{\rm 2D} \simeq 1.8\tau_3^{\rm 2D}$ under LO-phonon-dominated scattering, $\gamma_{\rm 2D}^{\rm app} \simeq 9.2$ eVÅ$^3$, consistent with spin relaxation in (001) GaAs quantum wells. This correlation-time renormalization clarifies why Dresselhaus coefficients extracted from two-dimensional spin relaxation can differ from the cubic bulk coefficient and provides a useful framework for interpreting linear and cubic Dresselhaus parameters in quantum wells.

## I. Introduction

Spin relaxation in GaAs has long served as a model problem for understanding spin dynamics in noncentrosymmetric semiconductors [1-3]. In zinc-blende semiconductors, the bulk inversion asymmetry gives rise to the Dresselhaus spin-orbit field, whose leading bulk contribution is cubic in the electron wave vector, $\mathbf{\Omega}_{\mathrm{D}}^{3\mathrm{D}} = 2\gamma_{3\mathrm{D}}/\hbar\left(k_x(k_y^2 - k_z^2), k_y(k_z^2 - k_x^2), k_z(k_x^2 - k_y^2)\right)$. The coefficient $\gamma_{3\mathrm{D}}$ is usually regarded as a material parameter characterizing the cubic Dresselhaus spin splitting in bulk GaAs. In D'yakonov-Perel' (DP) spin relaxation [4], however, the measured relaxation rate is not determined by the instantaneous spin splitting alone, but by the time-correlation integral of the spin-orbit field,

$$\Gamma_s = \frac{1}{\tau_s} \cong \int_0^\infty \langle \mathbf{\Omega}(t)\mathbf{\Omega}(0) \rangle dt.$$

Thus, the coefficient inferred from spin relaxation is always tied to the correlation kernel of the spin-orbit field. In (001)-oriented quantum wells, the conventional two-dimensional reduction of the Dresselhaus Hamiltonian replaces $k_z^2$ by the subband expectation value $\langle k_z^2 \rangle$. The leading in-plane Dresselhaus field then becomes linear in the in-plane wave vector, $\mathbf{\Omega}_{\mathrm{D}}^{2\mathrm{D}} \propto \gamma\langle k_z^2 \rangle (k_x, -k_y, 0)$, and the corresponding linear Dresselhaus coefficient is commonly written as $\beta_1 = -\gamma\langle k_z^2 \rangle$. This relation has been widely used to analyze spin-orbit coupling in GaAs/AlGaAs quantum wells. For example, Walser *et al.* measured the well-width dependence of $\beta_1$ in (001) GaAs quantum wells and extracted a bulk Dresselhaus coefficient of approximately $\gamma = -11 \pm 2$ eVÅ$^3$ using the standard relation between $\beta_1$ and $\langle k_z^2 \rangle$ [5]. Similar values were obtained in studies of persistent spin helix physics; Dettwiler *et al.* reported a Dresselhaus parameter of 11.6 $\pm 1$ eVÅ$^3$ in the analysis of stretchable persistent spin helices in GaAs quantum wells [6]. These studies demonstrate the usefulness of the projected two-dimensional Dresselhaus Hamiltonian for describing static or quasistatic spin-orbit couplings. Nevertheless, the same identification of the coefficient need not be automatic in a DP spin-relaxation measurement. The reason is that the projection $k_z^2 \to \langle k_z^2 \rangle$ does more than change the amplitude of the Dresselhaus field. It also changes the angular structure of the fluctuating field that enters the DP correlation integral. The original bulk Dresselhaus field is cubic in $\mathbf{k}$, and its relaxation is governed by a third-order angular correlation kernel, which we denote by $K_3^{3\mathrm{D}}$ . After projection onto a strict two-dimensional Hamiltonian, the leading field is linear in the in-plane momentum and is governed instead by a first-order momentum-correlation kernel, denoted by $K_1^{2\mathrm{D}}$. Therefore, the operation of projecting the Hamiltonian to two dimensions and the operation of evaluating the DP correlation kernel are not necessarily equivalent to retaining the original cubic field correlation and then taking a quasi-two-dimensional limit. This distinction becomes particularly important under room-temperature conditions where longitudinal optical (LO)-phonon scattering dominates the randomization of the spin-orbit field. In this regime, the first-order and third-order correlation kernels differ substantially. As we show below using nonballistic Monte Carlo simulations, the cubic Dresselhaus field in bulk GaAs is characterized by an effective correlation time of about $\tau_3^{3\mathrm{D}} \cong 150$ fs, and in the projected two-dimensional cubic Dresselhaus field in narrow (001) quantum wells $\tau_3^{2\mathrm{D}} \cong 130$ fs, whereas $\tau_{1+3}^{2\mathrm{D}} \cong 235$ fs for the first- and third-order hybridized kernel. Consequently, if the bulk cubic coefficient $\gamma_{3\mathrm{D}}$ is used unchanged in the projected two-dimensional DP model, the calculated spin relaxation rate is enhanced by approximately the ratio $\tau_{1+3}^{2\mathrm{D}}/\tau_3^{2\mathrm{D}}$. This leads to an overestimate of the spin relaxation rate in the two-dimensional regime. The central point of this work is that the coefficient appearing in a projected two-dimensional DP spin-relaxation model should be interpreted as an apparent coefficient associated with the projected correlation kernel, rather than as the intrinsic cubic bulk coefficient itself. If the same physical relaxation rate is written either in terms of the cubic-field correlation kernel or in terms of the projected two-dimensional kernel, $\Gamma_s = \gamma_{3\mathrm{D}}^2 \tau_3^{2\mathrm{D}} = \left(\gamma_{2\mathrm{D}}^{\mathrm{app}}\right)^2 \tau_{1+3}^{2\mathrm{D}}$ , then $\gamma_{2\mathrm{D}}^{\mathrm{app}} \cong \gamma_{3\mathrm{D}} \sqrt{\tau_3^{2\mathrm{D}}/\tau_{1+3}^{2\mathrm{D}}}$.

For LO-phonon-dominated scattering, our simulations give $\tau_{1+3}^{2\mathrm{D}} \cong 1.8\tau_3^{2\mathrm{D}}$, which naturally yields $\gamma_{2\mathrm{D}}^{\mathrm{app}} \cong \gamma_{3\mathrm{D}}/1.34$. Thus, a bulk value $\gamma_{3\mathrm{D}} \cong 12.4$ eVÅ$^3$ corresponds to an apparent two-dimensional DP coefficient $\gamma_{2D}^{app} \cong 9.2$ eVÅ$^3$. This value is close to the coefficient required to reproduce experimentally observed spin relaxation in (001) GaAs quantum wells [7]. In this paper, we test this interpretation by comparing bulk GaAs and (001) GaAs quantum wells within the same Monte Carlo framework. We first analyze bulk GaAs using inelastic nonballistic Monte Carlo simulations and show that the excitation-density dependence of the spin relaxation time is reproduced with $\gamma_{3D} \cong 12.4$ eVÅ$^3$ and a cubic-field correlation kernel $\tau_3^{3D} \cong 150$ fs. We then apply the same correlation analysis to (001) GaAs quantum wells in the room-temperature, weak-excitation limit. By calculating the well-width dependence of the effective correlation kernel, we show a continuous crossover from the bulk-like cubic-field value toward the two-dimensional linear-field value. Finally, we demonstrate that using the correlation-kernel-renormalized coefficient $\gamma_{2D}^{app}$, rather than the unchanged bulk coefficient, provides a consistent description of the observed spin relaxation in (001) quantum wells. Our results clarify why Dresselhaus coefficients inferred from different measurements in GaAs quantum wells can appear to differ. The intrinsic bulk coefficient $\gamma_{3D}$ should be associated with the original cubic Dresselhaus field, whereas the coefficient extracted from a projected two-dimensional DP spin-relaxation analysis is tied to the correlation kernel of the projected field. This distinction is also relevant for interpreting persistent spin helix experiments, where linear and cubic Dresselhaus terms are often compared using a common parameter $\gamma$, although their associated correlation kernels and experimental probes may be different.

## II. Monte Carlo model and correlation formalism

In this work we evaluate the spin-orbit-field correlation kernel and the corresponding D'yakonov-Perel' (DP) spin relaxation by a nonballistic ensemble Monte Carlo simulation [8]. The simulation is not based on conventional free-flight-time sampling. Instead, the carrier trajectory is advanced with a fixed small time $\Delta t$. At each time step, possible scattering events are tested by comparing random numbers with the corresponding scattering probabilities $W_i(\mathbf{k})\Delta t$, where $W_i(\mathbf{k})$ is the transition rate of the $i$-th scattering process. The time step is chosen sufficiently small so that the total scattering probability during one step satisfies

$$\sum_i W_i(\mathbf{k})\,\Delta t \ll 1.$$

If a scattering event occurs, its channel is selected according to the relative weights of the individual scattering rates. Otherwise, the electron wave vector and spin are propagated during the time interval $\Delta t$. The ensemble typically consists of $10^5$ electrons. The initial electron energy distribution is chosen according to the thermal equilibrium distribution at the lattice temperature. For the bulk GaAs simulations, an electron with an initial kinetic energy $E$ can undergo inelastic longitudinal-optical (LO) phonon absorption and emission processes, so that the electron energy changes by integer multiples of the LO-phonon energy, $E \to E \pm m\hbar\omega_0$ through successive LO-phonon scattering events. Here $\hbar\omega_0$ denotes the LO-phonon energy. The LO-phonon scattering is treated explicitly as an inelastic process. Acoustic-phonon scattering and electron-electron scattering [9] are included in the bulk calculation as additional momentum-randomizing processes, but are approximated as elastic scattering channels within a relaxation-time treatment. This approximation is used to incorporate their main contribution to the randomization of the spin-orbit field while keeping the computational cost manageable. The Dresselhaus spin-orbit field is written in the form $\mathbf{\Omega}(\mathbf{k}) = (2\gamma/\hbar)\mathbf{F}(\mathbf{k})$ where $\gamma$ is the Dresselhaus coefficient and $\mathbf{F}(\mathbf{k})$ contains the wave-vector dependence of the spin-orbit field. Equivalently, in terms of the spin-orbit Hamiltonian

$$H_{\mathrm{SO}} = \frac{\hbar}{2}\mathbf{\Omega}(\mathbf{k}) \cdot \boldsymbol{\sigma},$$

we define the quantity $\mathbf{F}(t) = \hbar\mathbf{\Omega}(t)/2\gamma$ which is independent of $\gamma$. This definition allows us to separate the material coefficient from the correlation kernel. The normalized spin-orbit-field correlation function is then calculated as $C_{F(t)} = \langle \mathbf{F}(t) \cdot \mathbf{F}(0) \rangle / \langle \mathbf{F}(0) \cdot \mathbf{F}(0) \rangle$ where the brackets denote an ensemble average over Monte Carlo trajectories and initial states. For the analysis of spin relaxation, we use the unnormalized correlation kernel $K = \int_0^\infty \langle \mathbf{F}(t) \cdot \mathbf{F}(0) \rangle \, dt$. When the relaxation of a particular spin component is considered, only the transverse components of the effective field are included. For example, for relaxation of $S_z$, $K_z = \int_0^\infty \langle F_x(t) F_x(0) \rangle + \langle F_y(t) F_y(0) \rangle dt$. The corresponding DP relaxation rate is then expressed as $\Gamma_z = 1/\tau_{s,z} = (2\gamma/\hbar)^2 K_z$ up to the convention used for the definition of $\mathbf{F}$. In the following, we use this correlation-kernel representation to compare the cubic Dresselhaus field in bulk GaAs with the projected field in $(001)$ quantum wells. In parallel with the correlation-kernel analysis, we also calculate the spin dynamics directly by solving the Bloch equation along each Monte Carlo trajectory, $d\mathbf{S}(t)/dt = \mathbf{\Omega}(t) \times \mathbf{S}(t)$ using a fourth-order Runge-Kutta method. The ensemble-averaged spin polarization is obtained as

$$\langle S_z(t) \rangle = \left(\frac{1}{N}\right) \sum_{j=1}^{N} S_{z,i}(t)$$

and the spin relaxation time is extracted from the decay of $\langle S_z(t) \rangle$. The correlation-kernel calculation and the direct Bloch-equation calculation provide complementary checks on the same underlying spin-orbit-field dynamics. In the bulk calculation, the Dresselhaus coefficient $\gamma_{3\mathrm{D}}$ is treated as an adjustable parameter and is determined by comparison with the experimentally measured spin relaxation time of bulk GaAs reported in our previous work [10]. Once $\gamma_{3\mathrm{D}}$ is fixed from the bulk calculation, the same Monte Carlo framework is used to evaluate the correlation kernels in $(001)$ quantum wells. This makes it possible to distinguish the intrinsic cubic Dresselhaus coefficient from the apparent coefficient obtained when the system is described by a projected two-dimensional spin-orbit field.

It should be emphasized that $K$ depends on the form of the spin-orbit field used in the simulation. In bulk GaAs, $\mathbf{F}(\mathbf{k})$ is the cubic Dresselhaus field. In a projected two-dimensional quantum-well model, however, the replacement $k_z^2 \to \langle k_z^2 \rangle$ changes the angular structure of $\mathbf{F}$ from cubic to linear in the in-plane momentum. The distinction between these two correlation kernels is central to the analysis below.

## III. Bulk GaAs

We first apply the Monte Carlo correlation analysis to bulk GaAs. This part serves two purposes. First, it provides a benchmark for the nonballistic Monte Carlo method by comparing the calculated spin relaxation time with the experimentally measured excitation-density dependence in bulk GaAs. Second, it defines the cubic Dresselhaus coefficient $\gamma_{3\mathrm{D}}$ and the corresponding cubic-field correlation kernel $K_3^{3\mathrm{D}}$, which are later used as the reference for the analysis of (001) quantum wells. In bulk GaAs, the Dresselhaus field is given by the cubic form

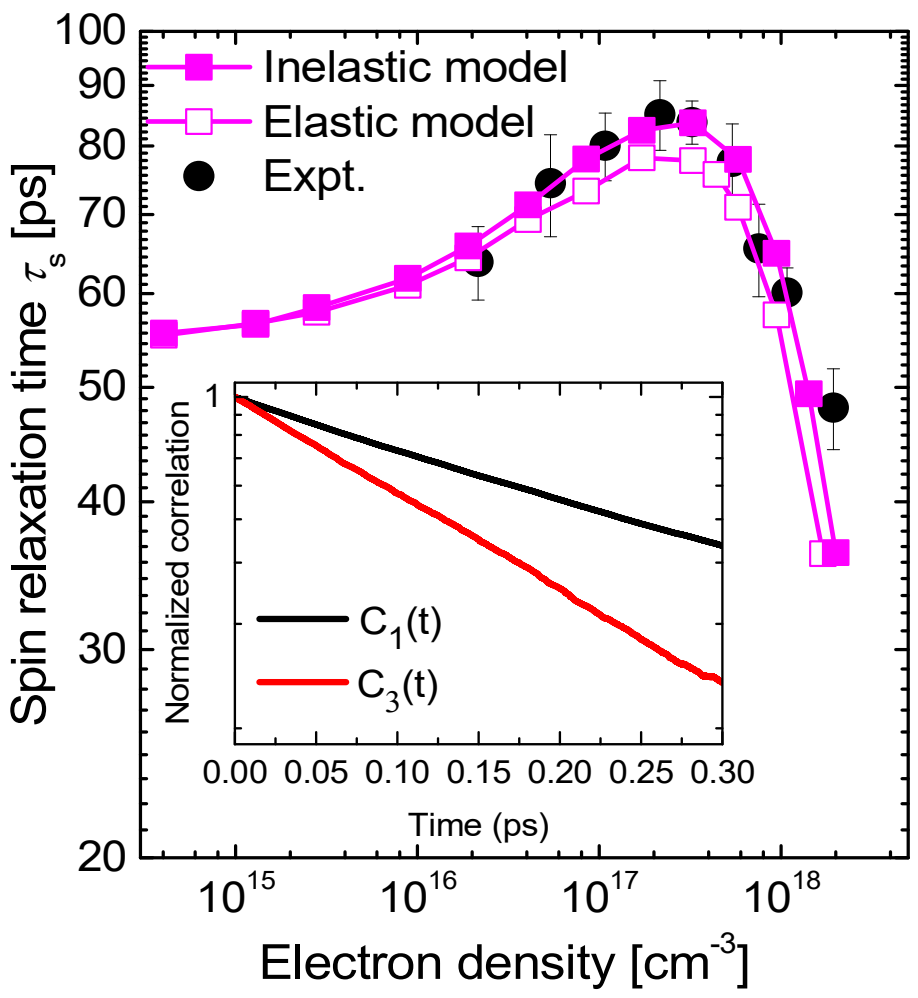

**Figure 1** Bulk GaAs benchmark of the nonballistic Monte Carlo simulation of the excitation-density dependence of the spin relaxation time in bulk GaAs at room temperature. Closed circles are experimental data from Ref. [10]; Closed squares are the inelastic Monte Carlo calculation including LO-phonon scattering, acoustic-phonon scattering, electron-electron scattering, and electron-hole scattering. The best agreement is obtained with $\gamma_{3\mathrm{D}} = 12.4$ eVÅ$^3$. Open squares are the results of Monte Carlo calculation based on the elastic scattering model, which reveals the best fit with $\gamma_{3\mathrm{D}} = 11.5$ eVÅ$^3$. The inset shows normalized correlation functions $C_3(t) = \langle \mathbf{F}_{3\mathrm{D}}(t) \cdot \mathbf{F}_{3\mathrm{D}}(0)\rangle / \langle \mathbf{F}_{3\mathrm{D}}(0) \cdot \mathbf{F}_{3\mathrm{D}}(0)\rangle$ of the cubic Dresselhaus field. The integrated correlation time is approximately 150 fs. The correlation function of the momentum, i.e. $C_1(t) = \langle \mathbf{k}(t) \cdot \mathbf{k}(0)\rangle / \langle \mathbf{k}(0) \cdot \mathbf{k}(0)\rangle$, is also shown. The integrated correlation is ~300 fs.

$$\mathbf{\Omega}_{\mathrm{D}}^{3\mathrm{D}}(\mathbf{k}) = \frac{2\gamma_{3\mathrm{D}}}{\hbar}\mathbf{F}_{3\mathrm{D}}(\mathbf{k}),$$

where $\mathbf{F}_{3\mathrm{D}}(\mathbf{k}) = \left(k_x\left(k_y^2 - k_z^2\right), k_y\left(k_z^2 - k_x^2\right), k_z\left(k_x^2 - k_y^2\right)\right)$. The spin-orbit-field correlation kernel relevant for the DP relaxation is then $K_3^{3\mathrm{D}} = \int_0^\infty \langle \mathbf{F}_{3\mathrm{D}}(t) \cdot \mathbf{F}_{3\mathrm{D}}(0)\rangle dt$. For an isotropic bulk system, the correlation of the cubic Dresselhaus field is governed primarily by a third-order angular correlation time. Therefore, $K_3^{3\mathrm{D}}$ is expected to be substantially shorter than the first-order momentum correlation time that appears in a projected two-dimensional linear Dresselhaus field. The bulk Monte Carlo simulation includes inelastic LO-phonon scattering explicitly. Acoustic-phonon scattering, electron-electron scattering, and electron-hole scattering are included as elastic momentum-randomizing processes within a relaxation-time approximation. The initial electron distribution is chosen to reproduce the experimental excitation conditions of our previous time-resolved spin-relaxation measurement in bulk GaAs [10]. For each excitation density, the electron distribution is equilibrated under the scattering processes, and the spin evolution is then calculated by solving the Bloch equation along the Monte Carlo trajectories, $d\mathbf{S}(t)/dt = \mathbf{\Omega}_{\mathrm{D}}^{3\mathrm{D}}(t) \times \mathbf{S}(t)$. The ensemble-averaged spin polarization $\langle S_z(t)\rangle$ is fitted by an exponential decay to obtain the spin relaxation time $\tau_{\mathrm{s}}$. Figure 1 compares the calculated spin relaxation time with the experimental excitation-density dependence of bulk GaAs [10]. The Dresselhaus coefficient $\gamma_{3\mathrm{D}}$ is treated as the only adjustable parameter in this comparison. The best agreement with experiment is obtained for $\gamma_{3\mathrm{D}} \cong 12.4$ eVÅ$^3$. This value is used as the reference cubic Dresselhaus coefficient in the following analysis. We emphasize that $\gamma_{3\mathrm{D}}$ is defined here as the coefficient of the original three-dimensional cubic Dresselhaus field. The calculated correlation function of the cubic Dresselhaus field is shown in the inset of Fig. 1. We plot the normalized correlation function $C_3^{3\mathrm{D}}(t) = \langle \mathbf{F}_{3\mathrm{D}}(t) \cdot \mathbf{F}_{3\mathrm{D}}(0)\rangle / \langle \mathbf{F}_{3\mathrm{D}}(0) \cdot \mathbf{F}_{3\mathrm{D}}(0)\rangle$. The correlation decays on a subpicosecond time scale. The effective correlation time, $\tau_3^{3\mathrm{D}} = \int_0^\infty C_3^{3\mathrm{D}}(t)dt$ is found to be approximately $\tau_3^{3\mathrm{D}} \cong 150$ fs under the present room-temperature, LO-phonon-dominated conditions. Equivalently, the unnormalized kernel $K_3^{3\mathrm{D}}$ extracted from the same trajectories provides the correlation factor that enters the DP relaxation rate, $\Gamma_s^{3D} = \left(\frac{2\gamma_{3D}}{\hbar}\right)^2 K_3^{3D}$. It is useful to compare this result with the corresponding calculation in an elastic-scattering approximation [10], as shown by open squares in Fig. 1. Although the details of the high-energy electron distribution and the short-time behavior of $C_3^{3D}(t)$ differ between the elastic and inelastic models, the effective cubic-field correlation kernel remains close to the value obtained above. Consequently, the Dresselhaus coefficient required to reproduce the bulk spin relaxation changes only weakly between the two models, remaining in the range $\gamma_{3D} \cong 12 \pm 0.5$ eVÅ$^3$. This stability indicates that the bulk spin relaxation provides a robust estimate of the cubic Dresselhaus coefficient within the present Monte Carlo framework. The important point for the following sections is that the bulk DP relaxation is characterized by a cubic-field correlation kernel with an effective correlation time of order $150$ fs. This kernel will be denoted by $K_3^{3D}$. In Sec. IV we show that the projected two-dimensional Dresselhaus field in $(001)$ quantum wells is governed instead by a first-order in-plane momentum correlation kernel, which is approximately twice as long in the strict two-dimensional limit. This difference in correlation kernels is the origin of the apparent reduction of the Dresselhaus coefficient inferred from two-dimensional spin relaxation.

## IV. Two-dimensional quantum wells

We next apply the same correlation-kernel analysis to $(001)$-oriented GaAs quantum wells. In contrast to bulk GaAs, the conventional two-dimensional reduction of the Dresselhaus Hamiltonian replaces the operator $k_z^2$ by its subband expectation value [11]. This converts the leading Dresselhaus field from a cubic function of the three-dimensional wave vector into a term that is linear in the in-plane wave vector. The purpose of this section is to clarify how this projection modifies the DP correlation kernel and how the corresponding apparent Dresselhaus coefficient changes from the bulk value.

### A. Model description

We consider undoped symmetric (001) GaAs quantum wells at room-temperature in weak-excitation limit. In this regime the carrier density is sufficiently low that the dominant randomization mechanism of the spin-orbit field is LO-phonon scattering. Therefore, in the main quantum-well Monte Carlo simulation, we explicitly include inelastic intra- and intersubband LO-phonon scattering [12]. Other processes, such as acoustic-phonon scattering and electron-electron scattering, can be incorporated in an elastic relaxation-time approximation as consistency checks, but are not included in the primary inelastic quantum-well calculation discussed below. For a (001) quantum well, the bulk Dresselhaus field is projected onto the quantum-well subbands. Starting from the cubic Dresselhaus form, $\mathbf{F}_{3\mathrm{D}}(\mathbf{k}) = \left(k_x\left(k_y^2 - k_z^2\right), k_y\left(k_z^2 - k_x^2\right), k_z\left(k_x^2 - k_y^2\right)\right)$ the conventional subband projection replaces the operator $k_z^2$ by its expectation value in the occupied subband, $k_z^2 \to \langle k_z^2 \rangle_n$ where $n$ is the subband index. The projected in-plane Dresselhaus field in subband $n$ is then written as $\mathbf{F}_{2\mathrm{D},n}(\mathbf{k}_\parallel) = \left(k_x(k_y^2 - \langle k_z^2 \rangle_n), k_y(\langle k_z^2 \rangle_n - k_x^2)\right)$, where $\mathbf{k}_\parallel = \left(k_x, k_y\right) = k_\parallel(\cos\phi, \sin\phi)$. Equivalently, $\mathbf{F}_{2\mathrm{D},n} = \mathbf{F}_1^{2\mathrm{D}} + \mathbf{F}_3^{2\mathrm{D}}$ with $\mathbf{F}_1^{2\mathrm{D}} = k_\parallel\left(\langle k_z^2 \rangle_n - \frac{k_\parallel^2}{4}\right)(-\cos\phi, \sin\phi)$, and $\mathbf{F}_3^{2\mathrm{D}} = -\frac{k_\parallel^3}{4}(\cos 3\phi, \sin 3\phi)$. Thus, although the leading term in narrow wells is the linear Dresselhaus contribution proportional to $\langle k_z^2 \rangle_n k_\parallel$, the Monte Carlo calculation retains both the projected linear term and the in-plane cubic Dresselhaus term. The correlation functions discussed below are calculated from the total projected field $\mathbf{F}_{2\mathrm{D},n}$, not from the linear term alone. The corresponding effective spin-orbit field is $\mathbf{\Omega}_{2\mathrm{D,n}}(\mathbf{k}_\parallel) = (2\gamma/\hbar)\ \mathbf{F}_{\mathbf{2D},\boldsymbol{n}}(\mathbf{k}_\parallel)$. In the conventional projected two-dimensional model, the same symbol $\gamma$ is often used as in the bulk cubic Dresselhaus Hamiltonian. However, as discussed in Sec. I, the projection changes the correlation kernel entering the DP relaxation. Therefore, in the following, we distinguish the cubic bulk coefficient $\gamma_{3\mathrm{D}}$ from the apparent coefficient $\gamma_{2\mathrm{D}}^{\mathrm{app}}$ used in the projected two-dimensional DP model. The intra- and intersubband LO-phonon scattering rates are calculated for each initial state $(n, \mathbf{k}_\parallel)$. In an intrasubband event, the subband index is unchanged, $n \to n$, whereas the in-plane kinetic energy changes by $\pm\hbar\omega_0$, subject to energy conservation. In an intersubband event, $n \to n'$, both the subband energy and the in-plane kinetic energy change according to $E_n + \frac{\hbar^2 k_\parallel^2}{2m^*} \to E_{n'} + \frac{\hbar^2 {k'}_\parallel^2}{2m^*} \pm \hbar\omega_0$. After each scattering event, the projected Dresselhaus field is recalculated using the new subband index and in-plane wave vector. In particular, intersubband scattering changes the value of $\langle k_z^2 \rangle_n$ entering the linear part of the projected Dresselhaus field. For each Monte Carlo trajectory, the subband index $n(t)$, the in-plane wave vector $\mathbf{k}_\parallel(t)$, and the total projected Dresselhaus field $\mathbf{F}_{2\mathrm{D}}(\boldsymbol{t}) = \mathbf{F}_{2\mathrm{D},n(t)}(\mathbf{k}_\parallel(t))$ are recorded. The spin-orbit-field correlation kernel relevant to $S_z$ relaxation is then computed as $K_{1+3}^{2\mathrm{D}}(L_\mathrm{w}) = \int_0^\infty \left( \langle F_x(t)F_x(0) \rangle + \langle F_y(t)F_y(0) \rangle \right) dt$, where $L_\mathrm{w}$ is the quantum-well width and $(F_x, F_y)$ are the components of the total projected field, including both the linear and cubic contributions. Explicitly, this kernel contains the linear-linear, cubic-cubic, and linear-cubic cross correlations: $K^{2\mathrm{D}} = K_{\mathrm{lin-li}} + K_{\mathrm{cub-c}} + K_{\mathrm{cross}}$. The normalized correlation function is defined as

$$C_{2\mathrm{D}}(t) = \frac{\langle F_x(t)F_x(0) \rangle + \langle F_y(t)F_y(0) \rangle}{\langle F_x^2(0) + F_y^2(0) \rangle}.$$

The corresponding effective correlation time is $\tau_{1+3}^{2\mathrm{D}}(L_\mathrm{w}) = \int_0^\infty C_{2\mathrm{D}}(t)dt$. It should be emphasized that $K_{1+3}^{2\mathrm{D}}(L_\mathrm{w})$ is the correlation kernel of the total projected two-dimensional Dresselhaus field. In the narrow-well, low-density limit, this kernel is dominated by the projected linear term and is therefore governed primarily by the first-order in-plane momentum correlation time $\tau_1^{2\mathrm{D}}$. However, the in-plane cubic Dresselhaus contribution is retained throughout the calculation. This projected-field kernel is different from the cubic-field kernel $K_3^{3\mathrm{D}}$ obtained in bulk GaAs, where the full three-dimensional cubic Dresselhaus field determines the DP correlation.

Although we use the notation $K_1^{2\mathrm{D}}$ for the strict two-dimensional limiting kernel in the following discussion, the actual Monte Carlo kernel $K_{1+3}^{2\mathrm{D}}(L_\mathrm{w})$ always includes the in-plane cubic Dresselhaus correction. The notation $K_1^{\mathrm{2D}}$ refers to the limiting behavior in which the projected linear term dominates the correlation.

### B. Well-width dependence of the correlation kernel: crossover between 3D and 2D

We now examine how the projected Dresselhaus-field correlation kernel evolves as a function of the lowest quantized energy $E_1$ by changing quantum-well width $L_{\mathrm{w}}$. Figure 2(a) shows the calculated correlation times $\tau_1^{\mathrm{2D}}$, $\tau_3^{\mathrm{2D}}$ and $\tau_{1+3}^{\mathrm{2D}}$ obtained from the normalized correlation function of the transverse Dresselhaus field. In wide wells, where the subband spacing becomes small and the system approaches the bulk-like limit, the correlation time approaches the cubic-field value obtained in bulk GaAs, $\tau_{1+3}^{\mathrm{2D}} \to \tau_3^{\mathrm{3D}} \cong 150$ fs as $E_1 \to 0$. In contrast, as the well width is reduced and the system approaches the strict two-dimensional limit, the projected Dresselhaus

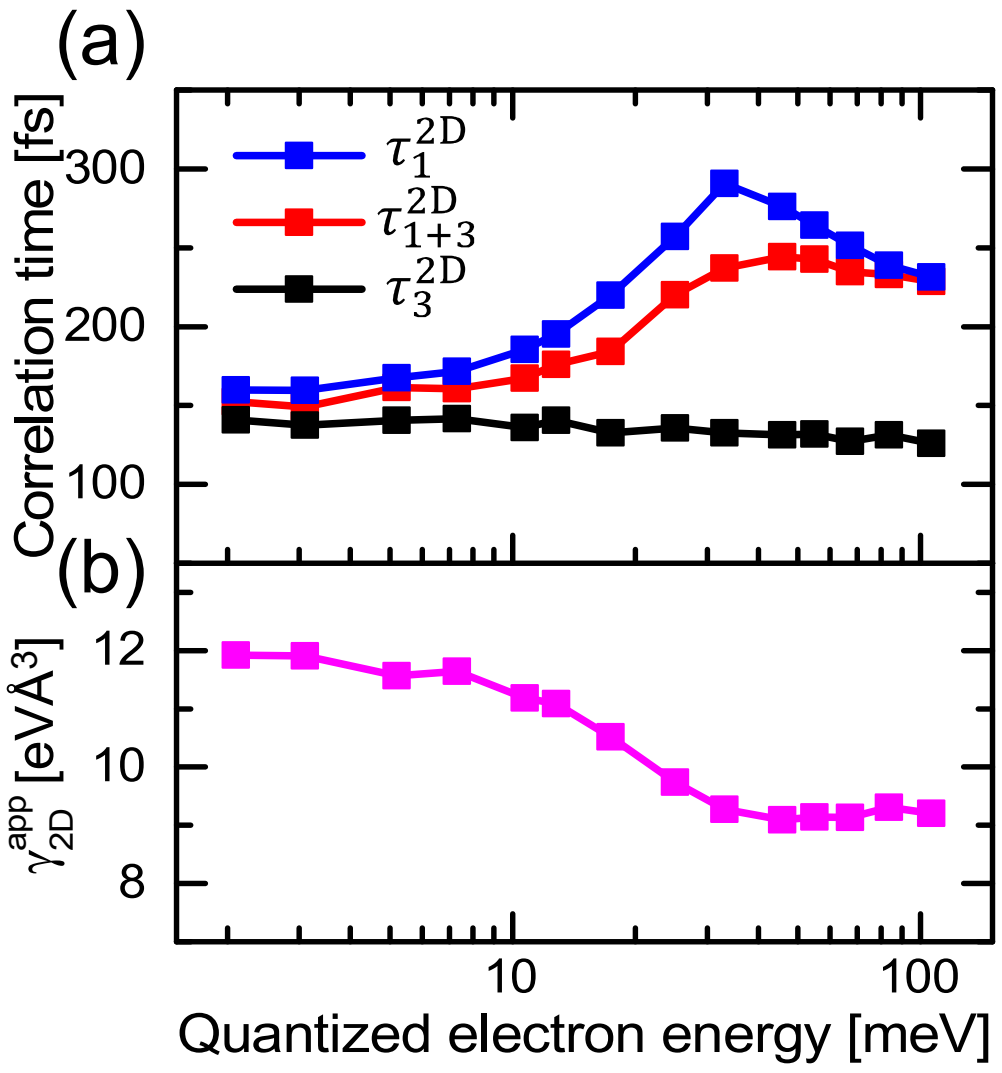

**Figure 2** (a) Quantized energy $E_1$ dependence of the projected Dresselhaus-field correlation times $\tau_1^{\mathrm{2D}}$ (blue), $\tau_3^{\mathrm{2D}}$ (black), and $\tau_{1+3}^{\mathrm{2D}}$ (red) in (001) GaAs quantum wells. The correlation times are obtained from $\tau_i^{\mathrm{2D}} = \int_0^\infty C_{\mathrm{2D},i}(t)dt$, where $C_{\mathrm{2D},i}(t)$ is the normalized correlation function of the transverse components $F_x$ and $F_y$. The crossover reflects the change of the DP correlation kernel from cubic to linear spin-orbit-field correlations. (b) $E_1$ dependence of $\gamma_{\mathrm{2D}}^{\mathrm{app}} = \gamma_{\mathrm{3D}}\sqrt{\tau_3^{\mathrm{2D}}/\tau_{1+3}^{\mathrm{2D}}}$.

field becomes dominated by the linear term proportional to $\langle k_z^2\rangle k_{\parallel}$. The correlation time then increases toward $\tau_{1+3}^{2D} \cong 235 \pm 5$ fs, while $\tau_3^{2D} \cong 130$ fs. This continuous evolution reflects the crossover of the relevant DP correlation kernel from the third-order angular correlation kernel of the cubic Dresselhaus field to the first-order in-plane momentum correlation kernel of the projected two-dimensional field. In the present LO-phonon-dominated regime, the first-order correlation time is approximately twice the third-order correlation time, $\tau_{1+3}^{2D} \cong 1.8\tau_3^{2D}$. This result is consistent with the expectation that higher-order angular harmonics are randomized more rapidly by momentum scattering than the first-order harmonic. At intermediate well widths, the calculated correlation time lies between these two limiting values. For example, around $L_{\mathrm{w}} \cong 15$ nm, the correlation time is typically about $184$ fs, whereas in the narrow-well limit it approaches $235$ fs. Thus, the Monte Carlo simulation directly shows a smooth correlation-kernel crossover, $K_3^{3D} \to K_{1+3}^{2D}(L_{\mathrm{w}}) \to K_1^{2D}$, as the quantum-well width is reduced. The important consequence is that the projected two-dimensional Hamiltonian does not merely change the magnitude of the spin-orbit field through $\langle k_z^2\rangle$. It also changes the time-correlation kernel entering the DP relaxation rate. Therefore, if the same bulk coefficient $\gamma_{3D}$ is used unchanged in the projected two-dimensional model, the increase of the correlation kernel from $K_3^{2D}$ to $K_1^{2D}$ directly enhances the calculated DP relaxation rate.

### C. Well-width dependence of spin relaxation time with the effective two-dimensional coefficient

We next discuss the consequence of the correlation-kernel crossover for their spin relaxation time. If the bulk cubic coefficient $\gamma_{3D} = 12.4$ eVÅ$^3$ is used unchanged in the projected two-dimensional Hamiltonian, the DP relaxation rate is $\Gamma_s^{\mathrm{fixed}}(L_{\mathrm{w}}) = \left(\frac{2\gamma_{3D}}{\hbar}\right)^2 K^{2D}(L_{\mathrm{w}})$. Because $K^{2D}(L_{\mathrm{w}})$ increases as the well width is reduced, this model predicts a pronounced shortening of the spin relaxation time in the two-dimensional regime. In particular, when the correlation kernel increases from the bulk-like value of approximately 150 fs to the two-dimensional value of approximately 235 fs, the calculated DP relaxation rate increases by roughly a factor of 1.5. Experimentally, however, such a strong reduction of the spin relaxation time is not observed in (001) GaAs quantum wells under the weak-excitation conditions considered here. This indicates that the coefficient used in the projected two-dimensional DP model should not be identified directly with the cubic bulk coefficient. Instead, the projected model should be written using an apparent coefficient $\gamma_{\mathrm{2D}}^{\mathrm{app}}$ defined by equating the relaxation rate written in terms of the cubic-field kernel and that written in terms of the projected two-dimensional kernel: $\gamma_{\mathrm{3D}}^{2}\tau_{3}^{\mathrm{2D}} = \left(\gamma_{\mathrm{2D}}^{\mathrm{app}}\right)^{2}\tau_{1+3}^{\mathrm{2D}}$. More generally, for an intermediate well width,

$$\gamma_{\mathrm{2D}}^{\mathrm{app}}(L_{\mathrm{w}}) = \gamma_{\mathrm{3D}}\sqrt{\tau_{3}^{\mathrm{2D}}(L_{\mathrm{w}})/\tau_{1+3}^{\mathrm{2D}}(L_{\mathrm{w}})}.$$

This expression yields a continuous crossover of the apparent coefficient from the bulk value to the two-dimensional value. In Fig. 2(b), $\gamma_{\mathrm{2D}}^{\mathrm{app}}$ is plotted as a function of $E_1$. In the wide-well limit, $K^{\mathrm{2D}}(L_{\mathrm{w}}) \to K_{3}^{\mathrm{3D}}$, and therefore $\gamma_{\mathrm{2D}}^{\mathrm{app}}(L_{\mathrm{w}}) \to \gamma_{\mathrm{3D}}$. In the strict two-dimensional limit, $K^{\mathrm{2D}}(L_{\mathrm{w}}) \to \tau_{1+3}^{\mathrm{2D}} \cong 1.8\tau_{3}^{\mathrm{2D}}$, and hence $\gamma_{\mathrm{2D}}^{\mathrm{app}}(L_{\mathrm{w}}) \to \gamma_{\mathrm{3D}}/1.34$. Using $\gamma_{\mathrm{3D}} = 12.4$ eVÅ$^3$, this gives $\gamma_{\mathrm{2D}}^{\mathrm{app}} \cong 9.2$ eVÅ$^3$. Figure 3 compares the calculated spin relaxation time with experimental data for (001) GaAs quantum wells [13-16]. The calculation using the fixed bulk coefficient $\gamma_{\mathrm{3D}} = 12.4$ eVÅ$^3$ predicts a substantial reduction of $\tau_s$ as the well width is decreased. In contrast, the calculation using the correlation-kernel-renormalized coefficient $\gamma_{\mathrm{2D}}^{\mathrm{app}}(L_{\mathrm{w}})$ gives a much smoother connection between the bulk-like and two-dimensional regimes and reproduces the experimental spin relaxation times more consistently. This result shows that the apparent coefficient inferred from two-dimensional DP spin relaxation is not a direct measure of the intrinsic cubic bulk coefficient. Rather, it is the coefficient appropriate for the projected two-dimensional spin-orbit field and its associated correlation kernel. The distinction between $\gamma_{3\mathrm{D}}$ and $\gamma_{2\mathrm{D}}^{\mathrm{app}}$ is therefore essential when comparing spin relaxation in bulk GaAs and (001) quantum wells.

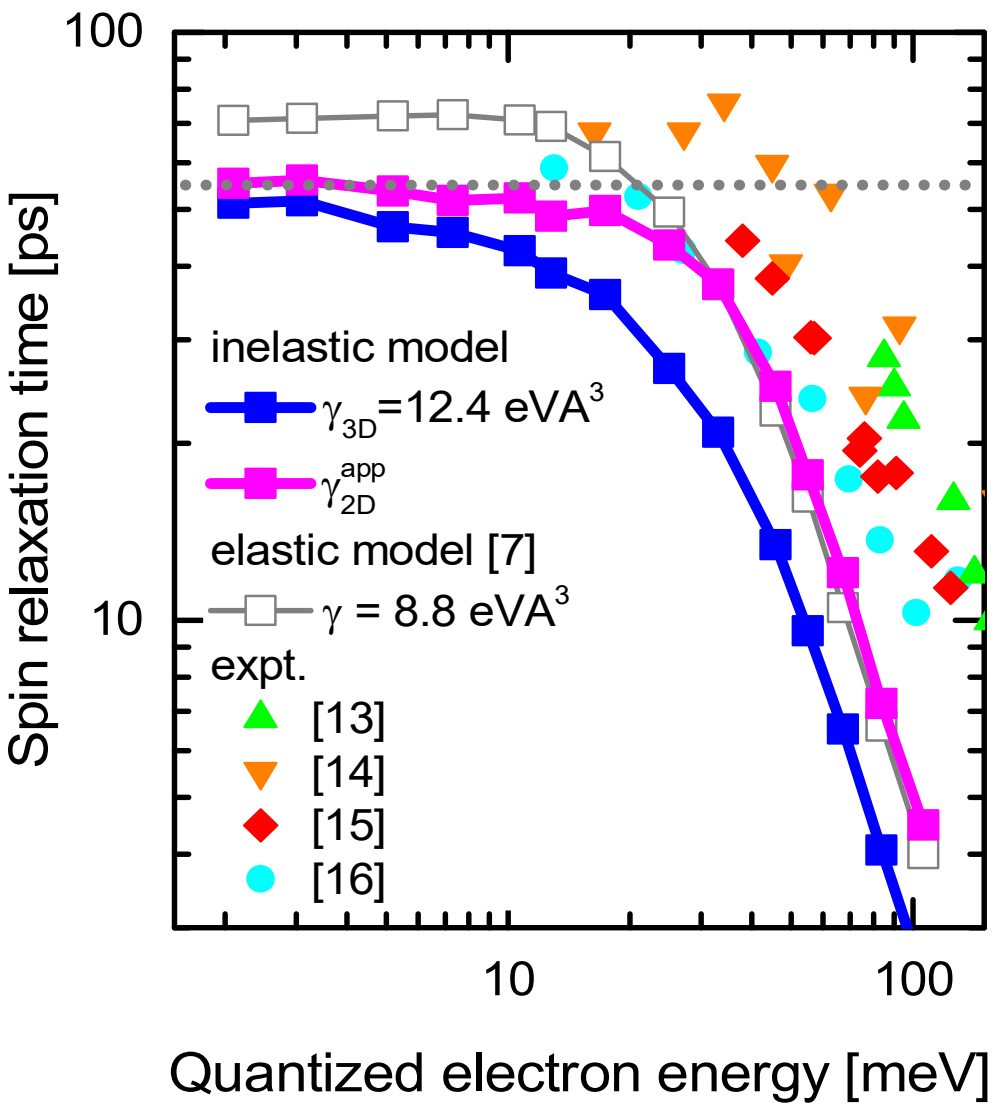

**Figure 3** Quantized energy dependence of the spin relaxation time in (001) GaAs quantum wells. The blue closed squares show the calculation obtained by using the fixed bulk cubic coefficient $\gamma_{3\mathrm{D}}$=12.4 eVÅ$^3$ in the projected two-dimensional Dresselhaus Hamiltonian. The purple solid squares show the result obtained using the correlation-time-renormalized coefficient $\gamma_{2\mathrm{D}}^{\mathrm{app}}(L_{\mathrm{w}})$. Gray open squares show the result obtained by elastic-scattering-model with $\gamma = 8.8$ eVÅ$^3$ [7]. A dotted line indicates the spin relaxation time of bulk GaAs. Other symbols are experimental spin relaxation times for (001) GaAs quantum wells [13-16].

## V. Discussion

The results presented above indicate that the Dresselhaus coefficient inferred from DP spin relaxation depends on the correlation kernel associated with the spin-orbit field used in the model. This does not imply that the intrinsic Dresselhaus coefficient of GaAs changes between bulk and quantum-well systems. Rather, it shows that the coefficient multiplying the projected two-dimensional field in a DP spin-relaxation analysis is an apparent coefficient tied to the projected correlation kernel.

### A. Distinction between $\boldsymbol{\gamma_{3D}}$ and $\boldsymbol{\gamma_{2D}^{app}}$

The intrinsic coefficient $\gamma_{3D}$ is naturally defined as the coefficient of the original cubic Dresselhaus field in bulk GaAs, $\mathbf{\Omega}_{\mathrm{D}}^{3\mathrm{D}} \propto \gamma_{3\mathrm{D}}\left(k_x(k_y^2-k_z^2), k_y(k_z^2-k_x^2), k_z(k_x^2-k_y^2)\right)$. When DP relaxation is evaluated using the correlation kernel of this cubic field, the relevant kernel is $K_3$, and the relaxation rate is written as $\Gamma_{\mathrm{s}}^{3\mathrm{D}} = (2/\hbar)^2\gamma_{3\mathrm{D}}^2 K_3^{3\mathrm{D}}$. In the bulk Monte Carlo calculation, this description gives $\gamma_{3\mathrm{D}} \cong 12.4$ eVÅ$^3$ and $\tau_3^{3\mathrm{D}} \cong 150$ fs. In contrast, in the conventional two-dimensional reduction of a (001) quantum well, one first replaces $k_z^2 \to \langle k_z^2\rangle$. The leading Dresselhaus field then becomes linear in the in-plane momentum: $\mathbf{\Omega}_{\mathrm{D}}^{2\mathrm{D}} \propto \gamma\langle k_z^2\rangle(k_x, -k_y, 0)$. The corresponding DP kernel is not the cubic-field kernel $K_3^{2\mathrm{D}}$, but the first-order in-plane momentum kernel $K_1^{2\mathrm{D}}$, or $K_{1+3}^{2\mathrm{D}}$. Therefore, if the projected two-dimensional model is used to describe the same spin relaxation physics, the coefficient in that model should satisfy ${\gamma_{2\mathrm{D}}^{\mathrm{app}}}^2 K_{1+3}^{2\mathrm{D}} = \gamma_{3\mathrm{D}}^2 K_3^{2\mathrm{D}}$. This gives $\gamma_{2\mathrm{D}}^{\mathrm{app}} = \gamma_{3\mathrm{D}}\sqrt{\tau_3^{2\mathrm{D}}/\tau_{1+3}^{2\mathrm{D}}}$. Under the room-temperature LO-phonon-dominated conditions considered here, the Monte Carlo calculation gives approximately $\tau_{1+3}^{2\mathrm{D}} \cong 1.8\tau_3^{2\mathrm{D}}$. Thus, $\gamma_{2\mathrm{D}}^{\mathrm{app}} \cong \gamma_{3\mathrm{D}}/1.34$. Using $\gamma_{3\mathrm{D}} = 12.4$ eVÅ$^3$, this yields $\gamma_{2\mathrm{D}}^{\mathrm{app}} \cong 9.2$ eVÅ$^3$. This relation explains why the projected two-dimensional DP model reproduces the spin relaxation data with a coefficient near 8.8 eVÅ$^3$ [7], whereas the bulk cubic-field analysis gives a coefficient near 12.4 eVÅ$^3$. The two values are not contradictory. They correspond to different correlation kernels: $K_3^{2\mathrm{D}}$ for the original cubic Dresselhaus field and $K_{1+3}^{2\mathrm{D}}$ for the projected two-dimensional linear and cubic field.

### B. Relation to the conventional $\boldsymbol{\beta_1 = -\gamma\langle k_z^2\rangle}$ description

The standard two-dimensional Dresselhaus Hamiltonian is commonly written in terms of the linear coefficient $\beta_1 = -\gamma\langle k_z^2\rangle$. This relation is appropriate for the static or quasistatic spin-orbit splitting of a single-subband two-dimensional electron system. It has been used successfully in experiments that measure the Dresselhaus spin-orbit coupling in GaAs quantum wells, including studies of well-width dependence and persistent spin helix physics [5,6]. The present result does not invalidate this static Hamiltonian description. Instead, it emphasizes that a DP spin-relaxation experiment measures the time-correlation integral of the effective spin-orbit field, not only the instantaneous spin splitting. Therefore, the coefficient extracted from spin relaxation depends on which spin-orbit field is used to define the correlation kernel. If the projected two-dimensional field is used, the relevant kernel is $K_{1+3}^{2\mathrm{D}}$, and the coefficient inferred from the relaxation rate is $\gamma_{2\mathrm{D}}^{\mathrm{app}}$. If the original cubic field is retained, the relevant kernel is $K_3^{3\mathrm{D}}$, and the corresponding coefficient is $\gamma_{3\mathrm{D}}$. In this sense, the apparent reduction of the coefficient in the projected two-dimensional DP analysis can equivalently be interpreted as a reduction of the effective confinement factor entering the dynamic spin-relaxation kernel. If one writes $\gamma_{2\mathrm{D}}^{\mathrm{app}}\langle k_z^2\rangle_{\mathrm{static}} = \gamma_{3\mathrm{D}}\langle k_z^2\rangle_{\mathrm{eff}}$, then $\langle k_z^2\rangle_{\mathrm{eff}}/\langle k_z^2\rangle_{\mathrm{static}} = \gamma_{2\mathrm{D}}^{\mathrm{app}}/\gamma_{3D} = \sqrt{\tau_3^{2\mathrm{D}}/\tau_{1+3}^{2\mathrm{D}}}$. For $\tau_{1+3}^{2\mathrm{D}} \cong 1.8\tau_3^{2\mathrm{D}}$, this ratio is approximately $1/1.34$. This does not mean that the static quantum-mechanical expectation value $\langle k_z^2\rangle_{\mathrm{static}}$ is incorrect. Rather, it means that the DP relaxation process probes a correlation-weighted effective quantity.

### C. Implications for comparing linear and cubic Dresselhaus terms

The distinction between $\gamma_{3\mathrm{D}}$ and $\gamma_{2\mathrm{D}}^{\mathrm{app}}$ is also relevant when comparing linear and cubic Dresselhaus terms in two-dimensional systems. In many analyses, both the linear term $\beta_1$ and the cubic Dresselhaus term are expressed using a common coefficient $\gamma$. This is natural at the level of the static Hamiltonian. However, in spin-relaxation measurements, the linear and cubic terms may be associated with different correlation kernels. For example, the projected linear term $\Omega_1 \propto \langle k_z^2 \rangle k_\parallel$ is governed primarily by a first-order in-plane momentum correlation. In contrast, a cubic Dresselhaus contribution retains higher angular harmonics and is randomized on a different time scale. Therefore, a coefficient extracted from the decay of a spin polarization or spin helix is not necessarily identical to a coefficient extracted from a static spin-orbit splitting measurement. This point is particularly important in persistent spin helix systems, where the balance between Rashba and renormalized Dresselhaus couplings is used to suppress spin relaxation. In such systems, $\beta_1$, cubic Dresselhaus corrections, and the corresponding spin lifetimes are often analyzed together. The present result suggests that care is required when interpreting an experimentally extracted $\gamma$, because the value may depend on whether the measurement probes a static spin-orbit field or a dynamical DP correlation kernel.

### D. Well-width crossover and physical interpretation

The well-width dependence of $K_{1+3}^{\mathrm{2D}}(L_{\mathrm{w}})$, or $\tau_{1+3}^{\mathrm{2D}}(L_{\mathrm{w}})$ provides a physical picture of the crossover between three-dimensional and two-dimensional spin relaxation. In wide wells, the subband spacing is small and the system retains more of the character of the original cubic Dresselhaus field. The correlation time approaches the bulk value, $\tau_{1+3}^{\mathrm{2D}}(L_{\mathrm{w}}) \to \tau_3^{\mathrm{3D}} \cong 150$ fs. In narrow wells, the single-subband two-dimensional description becomes increasingly appropriate. The projected field is dominated by the linear term proportional to $\langle k_z^2 \rangle k_\parallel$, and the correlation time approaches $\tau_{1+3}^{\mathrm{2D}}(L_{\mathrm{w}}) \to \tau_1^{\mathrm{2D}} \cong 235$ fs. If the coefficient is kept fixed at $\gamma_{\mathrm{3D}}$ throughout this crossover, the calculated spin relaxation rate necessarily follows the increase of $K_{1+3}^{\mathrm{2D}}(L_{\mathrm{w}})$. This would predict a pronounced shortening of the spin relaxation time in the two-dimensional regime. The available experimental data for (001) GaAs quantum wells do not support such a strong shortening. The smoother connection between bulk and quantum-well spin relaxation is obtained when the projected two-dimensional model is described using $\gamma^{\mathrm{app}}(L_{\mathrm{w}}) = \gamma_{\mathrm{3D}}\sqrt{\tau_3^{\mathrm{2D}}/\tau_{1+3}^{\mathrm{2D}}}$. This expression interpolates between $\gamma_{\mathrm{3D}}$ in the bulk-like limit and $\gamma_{3\mathrm{D}}/1.34$ in the strict two-dimensional limit. The crossover should not be interpreted as a change of the intrinsic material parameter. Rather, it reflects the change in the form of the effective spin-orbit field used to represent the system and the corresponding change in the DP correlation kernel.

### E. Validity of the elastic-scattering approximation

Before discussing the limitations of the present approach, we comment on the validity of the elastic-scattering approximation. Although the full inelastic Monte Carlo simulation is required for an absolute calibration of the Dresselhaus coefficient, the elastic approximation [7,10] captures the well-width dependence of the DP relaxation once the corresponding effective coefficient is properly calibrated.

In bulk GaAs, the elastic approximation gives a slightly smaller coefficient, $\gamma_{3\mathrm{D}}^{\mathrm{el}} \cong 11.5$ eVÅ$^3$, whereas the inelastic Monte Carlo simulation gives $\gamma_{3\mathrm{D}}^{\mathrm{inel}} \cong 12.4$ eVÅ$^3$. The ratio is $\gamma_{3\mathrm{D}}^{\mathrm{el}}/\gamma_{3\mathrm{D}}^{\mathrm{inel}} \cong 0.93$. A similar systematic difference is found in the quantum-well calculation. The full inelastic calculation gives an apparent two-dimensional coefficient of approximately $\gamma_{2\mathrm{D}}^{\mathrm{inel}} \cong 9.2$ eVÅ$^3$. Multiplying this value by the same bulk correction factor gives $9.2 \times \frac{11.5}{12.4} \cong 8.6$ eVÅ$^3$, which is close to the value $\gamma_{2\mathrm{D}}^{\mathrm{el}} \cong 8.8$ eVÅ$^3$ obtained in the elastic-scattering approximation. Moreover, the calculated well-width dependence of the spin relaxation time is nearly identical in the two approaches after this modest coefficient renormalization. This indicates that the elastic-scattering approximation slightly overestimates the spin relaxation rate because it neglects energy fluctuations associated with inelastic LO-phonon scattering, but it captures the essential correlation-kernel crossover and the well-width dependence of the DP relaxation. Therefore, the elastic approximation provides a useful and computationally efficient description once the corresponding effective coefficient is properly calibrated.

### F. Limitations

The present quantum-well calculation focuses on the room-temperature, weak-excitation regime, where LO-phonon scattering is the dominant process randomizing the spin-orbit field. This restriction is intentional, because it allows us to isolate the role of the Dresselhaus correlation kernel with minimal complications from high carrier density, electron-electron scattering, and electron-hole scattering. At higher excitation densities or at low temperatures, additional scattering processes can modify the correlation kernel and therefore the apparent coefficient inferred from spin relaxation. Moreover, the analysis presented here concerns spin relaxation in the projected two-dimensional DP model. It should not be interpreted as a direct measurement of the static Dresselhaus spin splitting. Independent measurements of $\beta_1$, for example based on spin-orbit-induced precession or persistent spin helix conditions, probe related but not identical quantities. A systematic comparison between static spin-orbit coupling measurements and DP correlation-kernel analyses would be valuable for clarifying the relation between $\gamma_{\mathrm{3D}}$ and experimentally extracted two-dimensional coefficients. Finally, the available experimental well-width dependence is still limited. A more complete experimental series covering the crossover region from wide, bulk-like wells to narrw, strictly two-dimensional wells would provide a direct test of the predicted relation $\gamma_{\mathrm{2D}}^{\mathrm{app}}(L_{\mathrm{w}}) = \gamma_{\mathrm{3D}}\sqrt{\tau_3^{\mathrm{2D}}/\tau_{1+3}^{\mathrm{2D}}}$. Nevertheless, the present results already provide a consistent explanation for why bulk GaAs is described by $\gamma_{\mathrm{3D}} \cong 12.4$ eVÅ$^3$, whereas $(001)$ quantum-well spin relaxation is described by an apparent coefficient close to $8.8$ eVÅ$^3$.

## VI. Conclusion

We have investigated the Dresselhaus coefficient inferred from D'yakonov-Perel' spin relaxation in bulk GaAs and $(001)$ GaAs quantum wells by using nonballistic Monte Carlo simulations. In bulk GaAs, the spin relaxation data are reproduced with a cubic Dresselhaus coefficient $\gamma_{3\mathrm{D}} \cong 12.4$ eVÅ$^3$. The corresponding spin-orbit-field correlation kernel is governed by the cubic Dresselhaus field and is characterized by an effective correlation time of approximately $\tau_3^{3\mathrm{D}} \cong 150$ fs. The extracted bulk coefficient is found to be relatively insensitive to whether the scattering model is treated in an elastic or inelastic manner, supporting its interpretation as the coefficient of the original three-dimensional cubic Dresselhaus field. In contrast, in a projected two-dimensional description of $(001)$ quantum wells, the replacement $k_z^2 \to \langle k_z^2 \rangle$ converts the leading Dresselhaus field into a term linear in the in-plane momentum. As a result, the DP correlation kernel changes from the cubic-field kernel $K_3^{3\mathrm{D}}$ to the first- and third-order hybridized in-plane momentum kernel $K_{1+3}^{2\mathrm{D}}$. Our quantum-well Monte Carlo simulations show that the effective correlation time evolves continuously from the bulk-like value of about $150$ fs in wide wells to approximately $235 \pm 5$ fs in the two-dimensional limit. This crossover reflects the change from cubic to projected linear spin-orbit-field correlations. Because the DP relaxation rate is proportional to the product of the squared Dresselhaus coefficient and the correlation kernel, the coefficient appearing in a projected two-dimensional DP model should not be identified directly with the cubic bulk coefficient. Instead, it should be interpreted as an apparent coefficient associated with the projected two-dimensional correlation kernel. Equating the relaxation rate expressed in terms of the cubic-field kernel and that expressed in terms of the projected two-dimensional kernel gives $\gamma_{2\mathrm{D}}^{\mathrm{app}} = \gamma_{3\mathrm{D}}\sqrt{\tau_3^{2\mathrm{D}}/\tau_{1+3}^{2\mathrm{D}}}$ For LO-phonon-dominated scattering, the simulations give $\tau_{1+3}^{2\mathrm{D}} \cong 1.8\tau_3^{2\mathrm{D}}$, and therefore $\gamma_{2\mathrm{D}}^{\mathrm{app}} \cong \gamma_{3D}/1.34 \cong 9.2$ eVÅ$^3$. This value provides a consistent description of spin relaxation in $(001)$ GaAs quantum wells, whereas using the bulk value $\gamma_{3\mathrm{D}}$ unchanged in the projected two-dimensional model overestimates the spin relaxation rate. These results show that the apparent Dresselhaus coefficient obtained from two-dimensional spin relaxation is not simply a direct measure of the intrinsic cubic bulk coefficient. Rather, it depends on the spin-orbit-field correlation kernel associated with the model used to describe the system. The intrinsic coefficient $\gamma_{3\mathrm{D}}$ should be associated with the original cubic Dresselhaus field, while $\gamma_{2\mathrm{D}}^{\mathrm{app}}$ is the effective coefficient appropriate for the projected two-dimensional DP kernel. This distinction provides a natural explanation for why bulk GaAs and (001) quantum wells can yield different apparent Dresselhaus coefficients without requiring a change in the underlying material parameter. The present analysis also suggests that care is required when comparing Dresselhaus parameters extracted from different experimental probes. Static or quasistatic measurements of spin-orbit splitting and dynamical measurements of DP spin relaxation need not return identical effective coefficients, because they are associated with different correlation kernels. This point is particularly relevant to quantum-well systems in which linear and cubic Dresselhaus terms are analyzed together, such as persistent-spin-helix structures. A systematic experimental study of the well-width dependence across the two-dimensional-to-bulk crossover would provide a direct test of the correlation-kernel renormalization proposed here.

This work was supported by JSPS KAKENHI (Grant Number JP24H00426, JP24K01391) and MEXT Initiative to Establish Next-generation Novel Integrated Circuits Centers (X-NICS) (Grant Number JPJ011438).